\documentclass{article}
\usepackage[latin9]{inputenc}
\usepackage{spconf,amsmath,graphicx}
\usepackage{amssymb,amsmath,epsfig}
\usepackage{lipsum,color,multirow,url}
\usepackage{arydshln, algorithm, algpseudocode}
\usepackage{microtype}
\usepackage{nth}
\usepackage[inline]{enumitem}
\usepackage{enumitem}
\usepackage{enumerate}
\usepackage{romannum}
\usepackage{textcomp}
\usepackage{comment}
\usepackage{float}

\title{Learning a Dual-Mode Speech Recognition Model via Self-Pruning}
\makeatletter
\def\name#1{\gdef\@name{#1\\}}
\name{\em Chunxi Liu,  Yuan Shangguan, Haichuan Yang, Yangyang Shi,  \\  \quad  \em Raghuraman Krishnamoorthi, Ozlem Kalinli}
\address{Meta AI, USA}
\begin{document}
\ninept
\maketitle
\begin{abstract}
There is growing interest in unifying the streaming and full-context automatic speech recognition (ASR) networks into a single end-to-end ASR model to simplify the model training and deployment for both use cases. 
While in real-world ASR applications, the streaming ASR models  typically operate under more storage and computational constraints - e.g., on embedded devices - than any server-side full-context models. Motivated by the recent progress in Omni-sparsity supernet training, where multiple subnetworks are jointly optimized in one single model, this work aims to jointly learn a compact sparse on-device streaming ASR model, and a large dense server non-streaming model, in a single supernet. 
Next, we present that, performing supernet training on both wav2vec 2.0 self-supervised learning and supervised ASR fine-tuning can not only substantially improve the large non-streaming model as shown in prior works, and also be able to improve the compact sparse streaming model. 
\end{abstract}
% One typical streaming ASR application is for on-device modeling, which
%
%
% ---------------------------------------------------------------------------------------------------------------------------------------------
\section{Introduction}
\label{sec:intro}

Thus far, end-to-end automatic speech recognition (ASR) models, which use neural networks to transduce audio into word sequences, have demonstrated state-of-the-art results compared to conventional hybrid speech recognizers. Specifically, recurrent neural network transducer (RNN-T) originally presented in \cite{graves2012sequence} has shown competitive ASR performance on various benchmarks \cite{chiu2019comparison, li2020comparison, zhang2021benchmarking}. Typically based on token emission latency, we categorize ASR models into: 
(i) streaming recognizers \cite{sainath2020streaming, mahadeokar2021flexi} that emit hypothesized words in real time, with low latency measured by milliseconds, and (ii) non-streaming models \cite{gulati2020conformer, zhang2020pushing} that only emit word hypotheses after processing the complete speech utterance. Latest streaming recognizers often employ a transformer/conformer encoder \cite{zhang2020transformer, li2021better}, and may use a limited future audio context (also referred to as look-ahead audio frames) \cite{shi2021emformer, shi2022streaming}.  Non-streaming recognizer takes the entire speech utterance as input, and scaling up the model size can often improve the model accuracies  \cite{zhang2020pushing}.  

Recently it has been shown favorable to unify the streaming and non-streaming models, either through a single shared encoder \cite{zhang2020transformer, yu2020dual, yao2021wenet, kim2021multi, weninger2022conformer}, or through cascaded streaming and non-streaming encoders \cite{li2021better, narayanan2021cascaded}. 
The efficacy of such unified or cascaded encoders includes that the previously two separate development and deployment workflows can be simplified into one process.
Note that in the two-pass cascaded encoders, input acoustic features are typically first processed by a streaming encoder, and a non-streaming encoder processes the streaming encoder outputs and aims to cover the first-pass accuracy loss. 
While for the unified dual-mode encoder, the non-streaming encoder directly processes the entire utterance and is immune from the accuracy degradation of the streaming encoder; additionally, the accuracy and latency of the streaming encoder can benefit from the weight sharing, or inplace knowledge distillation from the more performant non-streaming encoder \cite{yu2020dual}.

This work also focuses on the one-pass dual-mode encoder, while in practice, various streaming ASR models run on devices under more resource constraints, like disk size and memory footprint. In contrast, most non-streaming models run from the server with fewer constraints. Therefore, instead of developing equally sized encoders, it is preferable to jointly build a compact streaming model and a large non-streaming model for real-world ASR applications. We note that even though a single encoder is shared for both modes, we can substantially prune it into a featherweight, e.g., about 30M parameters as a streaming model, and use the original copy as a performant non-streaming encoder. Given the recent progress made in neural network pruning \cite{frankle2018lottery, wu2021dynamic, yang2022omni, ding2021audio}, we can specify a target sparsity level during model training, prune the model weights accordingly before inference, and finally obtain a model of the target model size. Meanwhile, we also aim to maintain the unpruned encoder's performance such that we can keep a copy of the original dense encoder and use it as a competitive non-streaming encoder.

Prior work \cite{yang2022omni} has shown success on the ASR training of varying sparsities jointly in a single model, also known as supernet training. A supernet is a shared-weight backbone network, where a subnetwork is extracted given each target sparsity level, and all the subnetworks are jointly optimized during supernet training. While it can facilitate ASR training of various model sizes, each sub-model in \cite{yang2022omni} operates with the same inference latency. 
Instead, this work focuses on two sparsity levels and two latency conditions: a high sparsity and low latency for the streaming model, and a zero sparsity (i.e., dense or unpruned) and full-utterance latency for the non-streaming model. 
Thus, in this case, the dual modes refer to the pruned/sparse streaming mode and the other unpruned/dense non-streaming mode.

Next, it has been widely shown that the self-supervised acoustic model pre-training based on wav2vec 2.0 \cite{baevski2020wav2vec} can substantially improve large non-streaming models; given sufficient unlabeled data, the potential accuracy gain can be proportional to the growing model size \cite{zhang2020pushing}. Similarly, achieving accuracy gains from  pre-training will be difficult given a compact model size. 
Also, very few works \cite{sainath2022improving} have shown the self-supervised pre-training efficacy in streaming models. In this paper, we present that by doing the dual-mode supernet training, self-supervised pre-training is not only able to substantially improve the large non-streaming model, and also to improve the compact sparse streaming model. 

% ---------------------------------------------------------------------------------------------------------------------------------------------
% ---------------------------------------------------------------------------------------------------------------------------------------------
\vspace{-0.2cm} 
\section{Supernet training of a dual-mode ASR Model}
\label{sec:supernet_asr}
\vspace{-0.2cm} 
% In this section we begin with a review of RNN-T based ASR, as originally presented in \cite{graves2012sequence}. Then we present our proposed 

% ---------------------------------------------------------------------------------------------------------------------------------------------
\subsection{RNN-T with Emformer encoder}

In this work we focus on the RNN-T based ASR models with the efficient memory transformer (Emformer) encoder \cite{shi2021emformer}. 

\subsubsection{RNN-T}

% ASR is formulated as a sequence-to-sequence problem. 
Each speech utterance is parameterized as an input acoustic feature vector sequence
$\textbf{x} = \{\textbf{x}_1 \ldots \textbf{x}_T\} = \textbf{x}_{1:T} $,  % \{\textbf{x}_t\}_{t=1}^T  
where $\textbf{x}_t \in \mathbb{R}^{d}$ and $T$ is the number of frames.  
Denote a grapheme set or a wordpiece inventory as $\mathcal{Y}$, and
the corresponding output sequence of length $U$ as $\textbf{y} = \{y_1 \ldots y_U\} = \textbf{y}_{1:U} $, where $y_u \in \mathcal{Y}$. 
We define $\bar{\mathcal{Y}}$ as $ \mathcal{Y} \cup \{ \emptyset \}$, where $\emptyset$ is the blank label.
Denote $\bar{\mathcal{Y}}^{*}$ as the set of all sequences over output space $\bar{\mathcal{Y}}$, and the element $\textbf{a} \in  \bar{\mathcal{Y}}^*$ as an alignment sequence.
% of length $\bar{U}$ represents the alignment between each time $t$ and an output label  $y_{\bar{U}}$ in $\bar{\mathcal{Y}}$.   factorize
Then we have the posterior probability:
\begin{equation}
    P( \textbf{y}  | \textbf{x}) =  \sum\limits_{ \textbf{a}  \in \mathcal{B}^{-1}(\textbf{y} ) }    P( \textbf{a}  | \textbf{x})
\label{eq:posterior}
\end{equation}
\noindent where $\mathcal{B}: \bar{\mathcal{Y}}^* \rightarrow  \mathcal{Y}^{*}  $ is a function that removes blank symbols from an alignment \textbf{a}.
A RNN-T model, $f(\textbf{x}; \theta)$, parameterizes the alignment probability $P(\textbf{a} | \textbf{x})$  with an encoder, a prediction network (predictor) and a joint network. 
The encoder $f^{\text{enc}}$ performs a mapping operation that converts $\textbf{x}$ into another sequence of representations 
$\textbf{h}^{\text{enc}}_{1:T} = \{\textbf{h}_1^{\text{enc}} \ldots \textbf{h}^{\text{enc}}_{T}\}$: 
\begin{equation}
 \textbf{h}^{\text{enc}}_{1:T} = f^{\text{enc}}(\textbf{x}; \theta^{\text{enc}}) 
 \label{eq:encoder}
\end{equation}
\noindent
A  prediction network $f^{\text{pred}}$ is to produce the new representation $\textbf{h}^{\text{pred}}_u$: 
\begin{equation}
    \textbf{h}^{\text{pred}}_{1:u} = f^{\text{pred}}(y_{0:(u-1)}; \theta^{\text{pred}})
\end{equation}
\noindent where $u$ is output label index and $y_0 = \emptyset$.
The joint network $f^{\text{join}}$  combines encoder output $\textbf{h}^{\text{enc}}_t$ and prediction network output $\textbf{h}^{\text{pred}}_u$ to compute logits $\textbf{z}_{t,u}$:
\begin{equation}
    \textbf{z}_{t,u} = f^{\text{join}}(\textbf{h}^{\text{enc}}_t, \textbf{h}^{\text{pred}}_u; \theta^{\text{join}}) 
\end{equation}
\begin{equation}
\begin{split} 
    P(y_u|  \textbf{x}_{1:t},  y_{1:(u-1)}) = \text{Softmax}(\textbf{z}_{t,u})   % \textbf{x}_1 \ldots \textbf{x}_t    y_1 \ldots y_{u-1} 
% & = P(y_u| \textbf{h}^{\text{enc}}_t, \textbf{h}^{\text{pre}}_u   ) \\            
\end{split}
\label{eq:posterior_1}
\end{equation}
\noindent such that the logits go through a softmax function and produce a posterior distribution of the next output label $y_u$ over $\bar{\mathcal{Y}}$. 
Note that, the posterior distribution in Eq. \ref{eq:posterior_1} is written as 
$P(y_u|\textbf{x}_{1:T}, y_{1:(u-1)})$, if it uses a non-streaming  encoder and takes each full-context utterance as inputs. 

% e.g., a full-context transformer network, 
% ---------------------------------------------------------------------------------------------------------------------------------------------
\subsubsection{Emformer encoder for streaming ASR} 
\label{ssec:emformer}

Chunk-based methods \cite{chen2021developing, yao2021wenet} have been widely applied for streaming ASR, and in this work, we use the block processing method with transformer encoder layers \cite{shi2021emformer}. The block processing chunks each whole utterance into a sequence of non-overlapping segments, $\textbf{x} = \{\textbf{C}_1 \ldots \textbf{C}_i \ldots \textbf{C}_I\}$, where $i$ is the index of a segment. 

To leverage the context information around each truncated segment, we concatenate a left contextual block $\textbf{L}_i$ (e.g., 20 acoustic frames or 120ms audio) and a respective right context block $\textbf{R}_i$  
(look-ahead context, e.g., 1 frame or 60ms) 
to each center block $\textbf{C}_i$,  to form a contextual segment $\hat{\textbf{C}}_i  = \{ \textbf{L}_i , \textbf{C}_i , \textbf{R}_i \}$. Then  during inference, a transformer encoder sequentially takes each $ \hat{\textbf{C}}_i$ as input, generates an output corresponding to each $\textbf{C}_i$, and forms a sequence of streaming outputs $\textbf{h}^{\text{enc}}_{1:t}$ (Eq. \ref{eq:encoder}).

%   we do not use the memory vector in the original Emformer layers.     and $\textbf{R}_i$   \begin{comment}  \end{comment}
% Finally, the RNN-T loss function is then the negative log posterior as in Eq. \ref{eq:posterior}:
% \begin{equation}  \mathcal{L}^{\text{\tiny{RNN-T}}} (\theta) = - \log  P(\textbf{y} |  \textbf{x} )  \end{equation}
% \noindent where $\theta$ denotes the model parameters.  is a feed-forward network that    takes both its state vector and the previous non-blank output label $y_{u-1}$ predicted by the model,
% Note that the encoder is analogous to an acoustic model, and the combination of prediction network and joint network can be seen as a decoder. 
% (i.e. transcription network in \cite{graves2012sequence})
% ---------------------------------------------------------------------------------------------------------------------------------------------
% ------------------------------------------------------------------------------------------------------------------------------------------------
\begin{figure}[t]
\centering
\includegraphics[width=0.9\linewidth,scale=0.9]{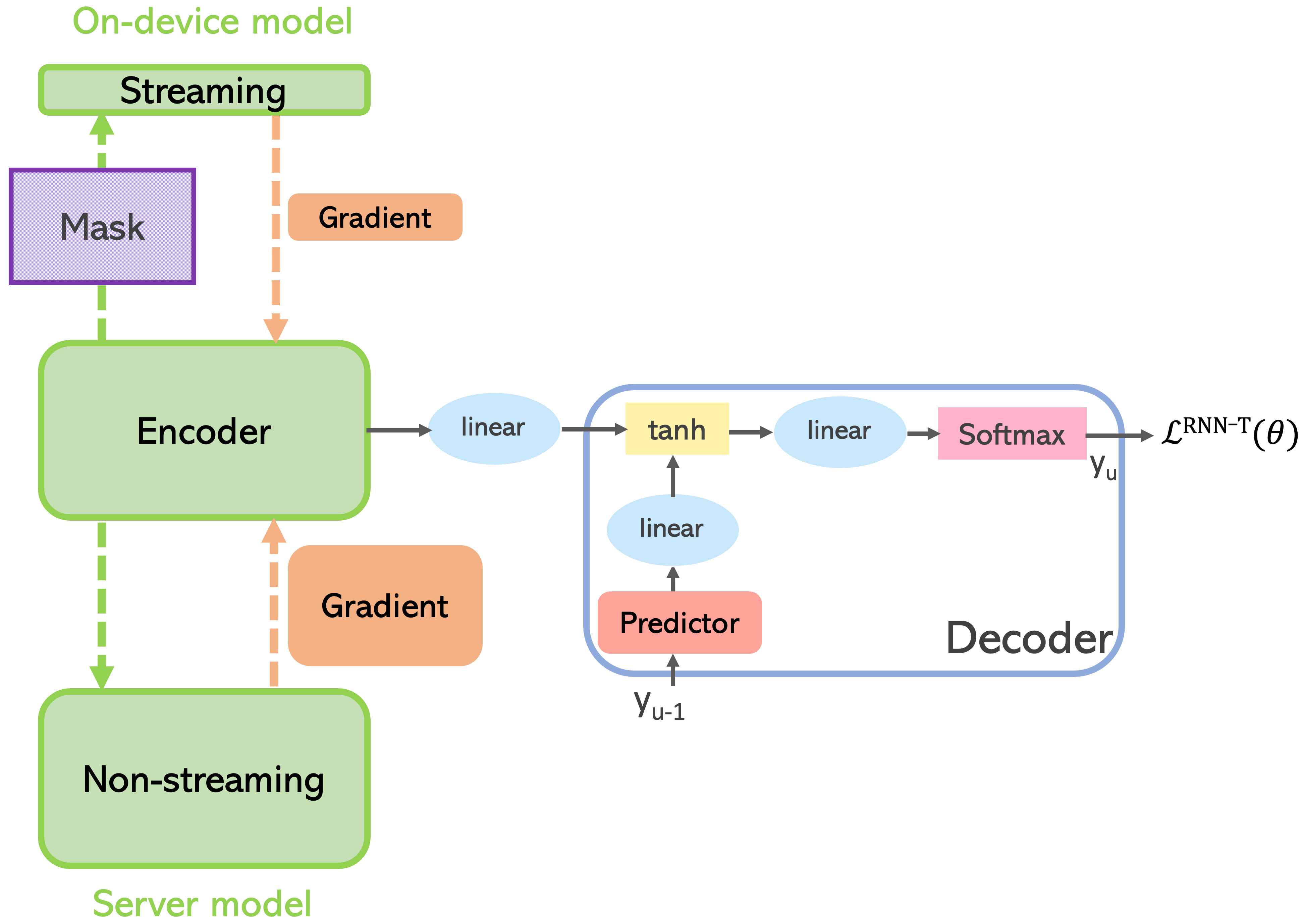}
\caption{\it Illustration of the proposed dual-mode ASR supernet training. When the encoder operates in the streaming mode, it is pruned by the binary mask (marked in purple). The predictor is pruned during streaming mode in the similar way, while intact during the non-streaming mode.}  %  In other words,
\label{fig:fig1}
\end{figure}
% ---------------------------------------------------------------------------------------------------------------------------------------------
% ---------------------------------------------------------------------------------------------------------------------------------------------
\subsection{Dual-mode ASR training via dynamic Emformer segment sampling}
\label{ssec:dual_mode}

As in Section \ref{ssec:emformer}, we note that the ASR latency depends on the length of the center block $\textbf{C}_i$, and changing the length of $\textbf{C}_i$ can effectively achieve the target latency. For example, when demanding an ultra-low latency, we can decrease $\textbf{C}_i$ to 100-200ms and use a minimal $\textbf{R}_i$ like 60ms or 0. Instead, to implement non-streaming ASR, we increase $\textbf{C}_i$ to a size as long as the full speech utterance and pad $\textbf{R}_i$ as 0. 

Thus to learn a dual-mode ASR model with both competitive streaming and non-streaming modes, at each training step, we randomly sample an Emformer segment length $|\textbf{C}_i|$, equally probable between a length of the target latency $\tau_0$ and a length equal to the longest utterance, $\tau_1$. 
Then the input utterances will be chunked differently based on the varying $|\textbf{C}_i|$. 
Both modes still use the same shared encoder, and only the query/key/value lengths vary according to $|\textbf{C}_i|$ in multi-head self-attention computations. The RNN-T decoder is also fully shared.  
This is similar to the domain-specific Emformer segment used in \cite{mahadeokar2021flexi}, where it applies a different segment length to each domain-specific data, though the models of different domains in \cite{mahadeokar2021flexi} are all low-latency streaming.

We implement it with the distributed data-parallel training across multiple GPUs \cite{ott2019fairseq}. Thus each GPU has a copy of the model, samples an $|\textbf{C}_i|$ between $\tau_0$ and $\tau_1$, and processes a sub-batch of data, after which gradients are then synchronized between GPUs for each model update, and the model learns both modes simultaneously.

% (all sub-batches constitute a mini-batch)
% ---------------------------------------------------------------------------------------------------------------------------------------------
\subsection{Dual-mode ASR supernet training}
\label{ssec:supernet}

As above, prior work \cite{zhang2020transformer, yu2020dual} and Section \ref{ssec:dual_mode} have described the joint training of a streaming and full-context model, in which both modes fully share the same parameters. 
Next, we aim to jointly learn a sparse streaming encoder and a dense full-context encoder. As in Figure \ref{fig:fig1}, during training both modes still share the same parameters, except that the pruning masks are only applied to the streaming mode. In this case it is a simplified supernet compared to \cite{yang2022omni}, as it contains only one performant sub-model for the streaming encoder.  

We denote a dense RNN-T model as $f(\textbf{x}; \theta)$, and a sub-model can be derived as $f(\textbf{x}; m \odot \theta )$ with a binary pruning mask $m \in \{0, 1\}^{|\theta|}$, where $\odot$ is the element-wise product. We perform layer-wise pruning \cite{yang2022omni} and prune the encoder Emformer layers and the predictor LSTM layer. 
A sparsity level $s$ denotes a percentage of the weights in each layer are pruned. We use an iterative magnitude pruning approach similar to  \cite{frankle2018lottery}, following the steps:
\begin{itemize}
%\begin{enumerate} %[(i)]
    \item[(i)] Training a unpruned dense model till a certain number of training updates $t_0$ (optionally with group lasso weight decay introduced in Section \ref{ssec:block_wd} below). As in Section \ref{ssec:dual_mode}, at each training step, we dynamically sample either a streaming or a non-streaming mode, and set the Emformer segment length $|\textbf{C}_i|$ accordingly. 
    \item[(ii)] Starting from $t_0$:
    \begin{itemize}
        \item[(a)] in each layer, for every $\Delta T$ training steps (i.e., pruning interval), prune $p$ (e.g., $p = 20\%$) of the weight parameters that have the smallest magnitudes. Pruning is done by setting the corresponding elements in the binary mask $m$ to $0$, and $m$ is updated every pruning interval $\Delta T$.
        \item[(b)] at each training step, when the streaming mode is sampled, the pruning masks are applied to the model weights during the forward-backward pass - gradients of the masked weights will be zero, and unmasked nonzero. When the non-streaming mode is sampled, pruning masks are not applied.
    %\end{enumerate}
    \end{itemize}
    \item[(iii)] After $n$ pruning intervals, i.e., $t_0 + n\Delta T $ training updates, $(1 - p)^n$ of weight parameters remain. Once the target sparsity level $s$ has been reached, $s = 1 - (1 - p)^n$,  the mask $m$ is not updated as in (ii, a) but fixed onward. The dual-mode training proceeds as in (ii, b).
%\end{enumerate}
\end{itemize}
\noindent Note that again the mode sampling (ii, b) is done on each GPU, and the gradients of each sub-batch are aggregated from all machines for each optimization step.
Also to obtain the sparsity speed-up from on-device hardware, all this work uses structured pruning, block size $8\times1$ as in \cite{yang2022omni}.

\section{Self-pruning via self-supervised learning}
\label{ssec:self_pruning}

% --------------------------------------------------------------------------------------------------------------------------------------
\subsection{Pre-training for ASR supernet}
\label{ssec:pretrain}

Prior works on self-supervised acoustic model pre-training is mostly focused on pre-training a non-streaming dense encoder with self-supervised criterion, and fine-tuning it with supervised ASR criterion. 
In this work we examine ways in which the encoder pre-training can improve the dual-mode ASR supernet, and the pruning masks learned during self-supervised training can be effective for the downstream ASR task.

We employ the wav2vec 2.0 pre-training criterion. During pre-training we either use a standard non-streaming encoder as in prior works \cite{baevski2020wav2vec, zhang2020pushing}, or use the dual-mode encoder as in Section \ref{ssec:dual_mode}, after which the pretrained model is fine-tuned  with RNN-T criterion, and then the encoder is always dual-mode to enable the dual-mode ASR.

Note that the encoder pruning, $(t_0, t_0 + n\Delta T)$ in Section \ref{ssec:supernet}, can be performed either during pre-training, or during RNN-T fine-tuning. In practice, we find pruning during RNN-T fine-tuning significantly underperforms pruning during pre-training. 
Note that the learning rate in RNN-T fine-tuning has to be small to maintain the pre-training effect, and we conjecture it is too small to adapt the encoder to the large sparsity changes. While the predictor is only used in RNN-T training, the LSTM layer is pruned during fine-tuning. 

\subsection{Pre-training with group lasso weight decay}
\label{ssec:block_wd}

Given sufficient unlabeled data, it can be helpful to prune from a converged model than pruning from scratch, so we consider increasing $t_0$ in Section \ref{ssec:supernet}. However, the model weights learned during the dense model training may not follow the $8\times1$ block structure as we use for the subsequent structured pruning, which results in performance degradation. Therefore, we particularly develop a block regularization technique below to fit the structured pruning.

In $8\times1$ block-wise pruning, essentially we would like the weights in each $8\times1$ block to be pruned or kept together. \emph{Group lasso}~\cite{yuan2006model} is a regularization method which selects grouped variables by penalizing the sum of $\ell_2$-norm of each group. In our case, we define each $8\times1$ block as a group, and specifically add a regularization term to the loss function $\mathcal{L}$:
\begin{equation}
    \min_{W} \mathcal{L} + \sum_{i=1}^l \lambda_i \sum_{g\in\mathcal{G}} \|W_g^{(i)}\|_2, \label{eq:block_lasso}
\end{equation}
where $l$ is the number of layers, $W_g^{(i)}$ is a certain $8\times1$ block in the $i$-th layer, and $\lambda_i$ is a hyper-parameter of penalty strength.
The subgradient with respect to $W_g^{(i)}$ in the block lasso term of Eq.~\ref{eq:block_lasso} is
\begin{equation}
{\lambda_i \over \|W_g^{(i)}\|_2} W_g^{(i)} \label{eq:block_lasso_grad},
\end{equation}
\noindent and the gradient descent direction pushes $W_g^{(i)}$ to zeros as weight decay, with strength $\lambda_i/\|W_g^{(i)}\|_2$. Thus the block regularization can push some weight blocks close to zeros, and keep other blocks almost unchanged.

As in many other regularizations, tuning $\lambda_i$ could be nontrivial. We propose to set it dynamically by the average value of the $\ell_2$-norm in $i$-th layer, i.e. $
\lambda_i = \lambda \sum_{g\in\mathcal{G}} \|W_g^{(i)}\|_2 / |\mathcal{G}|
$, where $\lambda$ is a global hyper-parameter shared for all layers, e.g., $\lambda=1$. 
In this way, we can greatly simplify the hyper-parameter tuning for such block regularization. 
Finally, we apply such group weight decay to the wav2vec 2.0 pre-training between  $(0, t_0 + n\Delta T)$ training updates, and turn it off afterwards. 

%\subsubsection{block lasso weight decay}
%\label{ssec:block_wd}
% ---------------------------------------------------------------------------------------------------------------------------------------
% ---------------------------------------------------------------------------------------------------------------------------------------------

\section{Experiments}
\label{sec:exp}

\subsection{Experimental setup}
\label{ssec:setup}

\subsubsection{Data}
\label{sssec:data}

We use the public LibriSpeech (LS) dataset \cite{panayotov2015librispeech} for all the supervised ASR experiments.
We apply speed perturbation \cite{ko2015audio} to the LS training data and produce three versions of each audio with speed factors $0.9$, $1.0$ and $1.1$.
We use the complete unlabeled Libri-Light dataset \cite{kahn2020libri} for self-supervised pre-training. We do not use the additional LibriSpeech language model (LM) corpus, and LM fusion is not applied in this work. 

\subsubsection{System implementation details}
\label{sssec:system}

Input acoustic features are 80-dimensional log-mel filterbank coefficients with 25 ms window size, and with mean and variance normalization. 
For all supervised ASR training, we use RNN-T criterion with alignment restrictions to improve training throughput \cite{mahadeokar2021alignment}, and apply the frequency and time masking as in SpecAugment \cite{park2019specaugment}.

RNN-T output labels consist of a blank label and 4096 wordpieces generated by the unigram language model algorithm from SentencePiece toolkit \cite{kudo2018sentencepiece}, and the joint network has 1024 hidden units, and a softmax layer of 4097 units. RNN-T predictor is a 1-layer LSTM of 512 hidden units, with dropout rate $0.3$. 

Six 80-dimensional log-mel features are concatenated with stride 6 to form a 480 dimensional vector, followed by a linear layer and mapped to an input to the encoder. For differing RNN-T model sizes, we vary the Emformer encoder parameters as in Table \ref{tab:encoder}. All encoders use relative positional embeddings with clipping distance 64 (3.84s) in self-attention \cite{shaw2018self}, dropout $0.1$, and the hybrid layer norm configurations 
\footnote{
Following \cite{wang2019transformer}, we find in each transformer layer, including the additional third layer norm  - that prevents the features from bypassing the transformer layer entirely - noticeably improves ASR accuracies.
}  \cite{wang2019transformer}.
Given the input feature stride 6, in streaming mode, Emformer left/center/right context lengths are 1.2s, 180ms, 60ms, i.e., $\textbf{L}_i=20, \textbf{C}_i =3, \textbf{R}_i =1$ (Section \ref{ssec:emformer}). In non-streaming mode, we set the center segment length as 36s, longer than any training utterance, to use the full context.

For all neural network implementation, we use an in-house extension of PyTorch-based \emph{fairseq} \cite{ott2019fairseq} toolkit. 
All experiments use multi-GPU  training, AdamW optimizer with decoupled weight decay 0.01 \cite{loshchilov2017decoupled}, $\beta_1=0.9$, $\beta_2=0.98$, and tri-stage \cite{park2019specaugment} learning rate schedule. 
The peak learning rate is 1e-3 for RNN-T training from scratch, 6e-4 for wav2vec 2.0 pre-training, tuned over $\{$2e-5, 5e-5$\}$  for RNN-T fine-tuning. For RNN-T, all ASR training uses global batch size 2560, up to 300 epochs on LibriSpeech. 
For wav2vec 2.0, pre-training on Libri-Light randomly crops each utterance into a max length of 15s on-the-fly, and the 181M dense models use global batch size 3072, for 300K training updates; 
since for supernet training, each training step has 50\% probability sampling the sparse sub-model on each GPU, where only a subset of the parameters have nonzero gradients, thus we use a larger global batch size 3840, and a longer training schedule of 400-450K updates.

As in Section \ref{ssec:supernet}, we prune all the encoder Emformer and predictor LSTM layers, with the following layer-wise sparsity level $s$ and pruning interval $\Delta T$: 
\begin{itemize}
\item $s=0.67, \Delta T=10K$ for training the 73M RNN-T model, 
\item $s=0.87, \Delta T=6K$  for training the 181M RNN-T model, 
\item $s=0.87, \Delta T=6K$ for pre-training the 181M model,
\end{itemize}
\noindent
such that the final sparse models after pruning have about 30M parameters in all cases. 
 In each pruning interval, we prune out $20\%$ remaining weights,  $p = 20\%$ as in \cite{ding2021audio}. 

% it randomly samples the initial masking time step with probability 0.065 
% efficient half precision floating point (FP16)  and epsilon $1e^{-8}$
% ----------------------------------------------------------------------------------------------------------------------------------------
% ----------------------------------------------------------------------------------------------------------------------------------------
% ---------------------------------------------------------------------------------------------------------------------------------------
\setlength{\tabcolsep}{0.16cm}
\begin{table}[H]
\caption{\label{tab:encoder} {\it  Emformer parameters for differing RNN-T model sizes.}}
\centering 
\begin{tabular}{   c | c  c c c c  c }
\hline \hline
    RNN-T    & \# layers  & embedding dim  & FFN dim & attn heads \\ 
\hline \hline
 35M     &  18      &     384  &   1024   &  4    \\
  73M     &  20      &     512  &   2048   &  8    \\    
   181M    &  24      &     768  &   3072   &  8    \\  
\hline \hline
\end{tabular}
\end{table}
% ----------------------------------------------------------------------------------------------------------------------------------------

% ----------------------------------------------------------------------------------------------------------------------------------------
\setlength{\tabcolsep}{0.01cm}
% \renewcommand{\arraystretch}{1.0}
%\begin{table*}[t]
\begin{table}[H]
\caption{\label{tab:result1} {\it  WER results on LibriSpeech test-other. For system \textbf{D2}, we use a non-streaming encoder for the first 50K updates, and then switch it to the dual-mode encoder afterwards and perform training the same as \textbf{D1}. } }
\centerline{ 
% \begin{tabular}{  p{4.01cm}  | c | p{0.8cm} |  p{1.98cm}  p{0.8cm} | c  }     In Section \ref{ssec:aux_loss} 
\begin{tabular}{ l | c  |   c   }
\hline \hline
     & unpruned/pruned   & unpruned \\ 
     & streaming  &  non-streaming   \\ 
\hline 
 \textbf{B1} \  35M, streaming dense         &  11.2    &   -   \\  
 \textbf{B2} \  35M, dual-mode dense         &  10.9    &  8.7  \\  \cdashline{1-3}[1.0pt/0.5pt]
 \textbf{C1} \  73M, streaming sparsity 0.67   &  10.9    &  -    \\  
 \textbf{C2} \  73M, non-streaming dense       &   -      &  6.4  \\  \cdashline{1-3}[1.0pt/0.5pt]
 \textbf{D1} \  73M, streaming sparsity 0.67,  &  \multirow{2}{*}{ 10.6 }  &  \multirow{2}{*}{ 7.0 }   \\
         non-streaming dense                   &                      &          \\ 
 \textbf{D2} \ non-streaming dense, 50K + \textbf{D1}  &    10.4      &     6.6   \\ 
\hline \hline
\end{tabular}}
\end{table}  
% ------------------------------------------------------------------------------------------------------------------------------------------
% --------------------------------------------------------------------------------------------------
\setlength{\tabcolsep}{0.127cm}
\begin{table*}[h]
\caption{\label{tab:result3}{\it  WER results of 181M dense models on LibriSpeech (LS) test sets. Pre-training randomly crops each utterance on-the-fly into max length 15s for system \textbf{B1} and \textbf{B2}, 30s for \textbf{B3}.   
All streaming ASR uses center context 180ms, right context 60ms, and 240ms latency in total (Section \ref{sssec:system}). LM fusion is not used.}}
\centerline{ 
% \begin{tabular}{  p{1.4cm}  p{2.6cm}  |  p{1.2cm}  p{1.2cm} p{1.2cm} p{1.2cm}   | p{1.2cm}   | p{1.2cm}  }
\begin{tabular}{ c | l   |    c c  |  c   c }
\hline \hline
\multirow{2}{*}{\textbf{dataset}} & \multirow{2}{*}{\textbf{system} }  & test-clean & test-other & test-clean & test-other  \\  \cdashline{3-6}[1.0pt/0.5pt] 
    &      &   \multicolumn{2}{c|}{unpruned streaming}      & \multicolumn{2}{c}{unpruned non-streaming}   \\  \hline \hline
 LS & \textbf{B0} \ 181M, dual-mode, dense  &  4.1    &   9.7  &  3.1 &    7.1  \\    \hline
\multirow{3}{*}{ Libri-Light + LS }  & \textbf{B1} \  dual-mode wav2vec, pretrain on 15s segment, dense + \textbf{B0}   &   3.3   &   8.3  & 2.3  &  5.3 \\
         & \textbf{B2} \  non-streaming wav2vec,  pretrain on 15s segment, dense + \textbf{B0}  & 3.1  &  7.8   &  2.1  &  4.3    \\   
         & \textbf{B3} \  non-streaming wav2vec,  pretrain on 30s segment, dense + \textbf{B0}  & 3.0  &  7.4   &  2.1  &  4.3    \\   
\hline \hline
\end{tabular}}
%\vspace{-0.2cm} 
\end{table*}

\setlength{\tabcolsep}{0.1cm}
\begin{table*}[h]
\caption{\label{tab:result4}{\it  WER results of 181M supernet models. Pre-training randomly crops each utterance into max length 15s in all systems below.
As Section \ref{ssec:pretrain}, supernet training refers to using sparsity 0.87 to learn a sparse sub-model, and using the unpruned model to learn a dense encoder.   All streaming ASR uses center context 180ms, right context 60ms, and has about 32M parameters after pruning. LM fusion is not used.}}
\centerline{ 
% \begin{tabular}{  p{1.4cm}  p{2.6cm}  |  p{1.2cm}  p{1.2cm} p{1.2cm} p{1.2cm}   | p{1.2cm}   | p{1.2cm}  }
\begin{tabular}{ c | l   |    c c  |  c   c }
\hline \hline
%    &   &   \multicolumn{2}{c|}{pruned streaming}   &  \multicolumn{2}{c}{unpruned non-streaming}    \\    \cdashline{3-6}[1.0pt/0.5pt]  
\multirow{2}{*}{\textbf{dataset}} & \multirow{2}{*}{\textbf{system} }  & test-clean & test-other & test-clean & test-other  \\  \cdashline{3-6}[1.0pt/0.5pt] 
    &      &   \multicolumn{2}{c|}{pruned streaming}      & \multicolumn{2}{c}{unpruned non-streaming}   \\  \hline \hline
 LS  & \textbf{C1} \quad 181M, streaming sparsity 0.87, non-streaming dense    &  4.4 &  11.2  & 2.7  &  6.4  \\  \hline
\multirow{5}{*}{Libri-Light + LS}   & \textbf{C2} \quad  non-streaming wav2vec supernet training, 400K updates + \textbf{C1} & 4.6  &  12.1  & 2.5  &  5.3    \\ \cdashline{2-6}[1.0pt/0.5pt]
&\textbf{C3.1} \ non-streaming wav2vec with group lasso, 50K updates + &\multirow{2}{*}{ 3.9 } &\multirow{2}{*}{ 10.2 } & \multirow{2}{*}{ 2.3 } & \multirow{2}{*}{ 4.5 } \\
 & \qquad \ \ non-streaming wav2vec  supernet training, 350K updates + \textbf{C1}   &    &  &  &   \\  \cdashline{2-6}[1.0pt/0.5pt] 
 &  \textbf{C3.2} \ non-streaming wav2vec with group lasso, 150K updates + & \multirow{2}{*}{ 3.9  } & \multirow{2}{*}{ 9.4 } & \multirow{2}{*}{ 2.2  }  & \multirow{2}{*}{ 4.3 } \\
 & \qquad \ \ non-streaming wav2vec  supernet training, 300K updates + \textbf{C1} &  & &   &  \\ % \cdashline{1-5}[1.0pt/0.5pt] 
%
% \bold{C4} \   \bold{C3}  + IPL                           &      &         &         &          \\
\hline \hline
\end{tabular}}
%\vspace{-0.2cm} 
\end{table*}
%\vspace{-0.2cm} 
% ----------------------------------------------------------------------------------------------------------------------------------------
% ----------------------------------------------------------------------------------------------------------------------------------------
% ----------------------------------------------------------------------------------------------------------------------------------------

\subsection{Results of the dual-mode ASR supernet}
\label{ssec:results_1}

We first build a pair of 35M dense model baselines: a streaming single-mode dense model B1, and a streaming and non-streaming dual-mode model B2.
ASR word error rate (WER) results on LibriSpeech \emph{test-other} are shown in Table \ref{tab:result1}.
We find B2 slightly outperforms B1, as observed in \cite{yu2020dual} similarly. 
Then we build a pair of 73M models:  
\begin{itemize}
\item[(i)]  a single-mode sparse streaming model C1 with sparsity  0.67, so after pruning it has about 29M parameters, less than B1 and B2, 
\item[(ii)]  a single-mode dense non-streaming model C2, 
\end{itemize}
\noindent
such that respectively, the separate single-mode C1 and C2 use the same number of parameters as the proposed dual-mode supernet model D1. 
We find the sparse streaming mode of D1 outperforms both dense models B1, B2 and the single-mode C1, but the D1 unpruned non-streaming mode falls behind C2. 

D1 uses $t_0 = \Delta T = 10K$ above (Section \ref{ssec:supernet}), and we find simply increasing $t_0$ is not helpful. 
Then we try a two-step approach in system D2:
\begin{itemize}
\item[1.] increase $t_0=50K$, and use a single-mode non-streaming encoder, i.e., always use the full context between $(0, t_0)$,
\item[2.] then after $t_0$, switch it to the dual-mode encoder, and perform training the same as D1. 
\end{itemize}
Then we find D2 to provide non-streaming performance on a par with C2. 
Overall, we demonstrate the efficacy of jointly learning a sparse streaming sub-model and a dense non-streaming model in a single supernet.

% We also find a noticeable use case of such dual-mode supernet training. Note that if we aim to build a compact streaming model with iterative pseudo-labeling (IPL), and the standard IPL will keep using the same streaming model for both pseudo-labeling and model training. However, the proposed dual-mode supernet training can effectively apply the unpruned non-streaming mode to pseudo-labeling, such that the resulting pseudo-label quality is much better than the ones from a compact streaming model. We briefly experiment with such method on Libri-Light, and we find system D2 can indeed provide large WER reductions for both modes. Ultimately we can both obtain a target sized streaming model after pruning, and without pruning, obtain a performant non-streaming model.  
% ---------------------------------------------------------------------------------------------------------------------------------------------
\subsection{Results of the  pre-training efficacy on dual-mode ASR }
\label{ssec:results_2}

Then we scale up the model size to 181M, as in Table \ref{tab:result3}\footnote{We note that, training the large transformer/Emformer models like system B0 from scratch - without additional regularization techniques - significantly underperforms. While applying auxiliary training criteria \cite{liu2021improving} would substantially improve baseline B0, they can be applied to other competing systems like B1 and B2 as well, so we will leave it to future work.}, and first examine the pre-training effects on dense models. 

As in Section \ref{ssec:pretrain},  we perform the wav2vec 2.0 pre-training on Libri-Light, and then afterwards use dual-mode encoder during RNN-T fine-tuning, to enable the dual-mode ASR.  We also try using the dual-mode encoder during wav2vec pre-training as well, referred to as the dual-mode wav2vec in B1 (see Table \ref{tab:result3}). However, by comparing B1 and B2, we find pre-training with just the non-streaming encoder instead is much more effective for both non-streaming and streaming ASR. Note that system B1 and B2 are pre-trained on audio segments cropped up to 15s, and we further increase the max segment length to 30s on system B3. We find B3 can produce further better streaming results compared to B2.

In all cases above, we present that pre-training can not only substantially improve the non-streaming ASR results as widely shown in prior works, and also noticeably improve streaming ASR performance, as one of the contributions in this work. The proposed dynamic Emformer segment sampling (Section \ref{ssec:dual_mode}) allows for using a non-streaming encoder to maximize the pre-training benefits, and enabling the high-performing dual-mode ASR afterwards.

\subsection{Results of supernet training with both self-supervised and supervised criteria} 
\label{ssec:results_3}

Next, as in Table \ref{tab:result4}, we first build a dual-mode supernet model C1 with labeled data only, and then start to use unlabeled data and examine the pre-training effects on both the sparse streaming mode and the dense non-streaming mode.

As discussed in Section \ref{ssec:pretrain}, we find any encoder pruning during RNN-T fine-tuning results in severe streaming ASR degradation, significantly falling behind the baseline C1. 
Thus instead we prune the encoder during pre-training.  
Note that for the ASR supernet training (Section \ref{ssec:supernet}), we will sample between streaming and non-streaming modes;  
however, given the result comparison between B1 and B2, we always use non-streaming mode during pre-training - we sample between the sub-model and the whole model (i.e., apply the mask or not), and both operate in the non-streaming mode.

Thus the encoder pruning mask is learned completely on the unlabeled data without supervision, and the encoder mask is fixed during RNN-T fine-tuning, so we refer to such process as self-pruning. The predictor is also pruned for streaming ASR, and the predictor mask is learned during RNN-T fine-tuning.   
Additionally, after such supernet training, the identified sparse sub-model will go through different post-processing and specialized hardware for storage and run-time optimization, therefore, we can choose separate best checkpoints across epochs for the sparse streaming sub-model and the dense non-streaming model respectively, based on the ASR accuracies on LS \emph{dev-other} subset. 

%  the sparse model is made by pruning a dense model  we will make the original copy

Following such training algorithm, although the system C2 gives higher non-streaming accuracies than the baseline C1 without pre-training, C2 still trails C1 on the streaming accuracy\footnote{Although by comparing the dense model B2 and B3 (Table \ref{tab:result3}),  pretraining on 30s audio segments is more effective for streaming ASR than on 15s, we find such observation does not hold true for the supernet training like system C3.1. We conjecture the explanation that pretraining on longer segments for a highly sparse model results in more difficult neural network optimization problems, e.g., the training will diverge using the same learning rate 6e-4, and we have to use 4e-4. Thus instead, system C2, C3.1 and C3.2 (Table \ref{tab:result4}) are all pre-trained on segments up to 15s.}. 
Then we note that C2 performs iterative pruning from scratch, i.e.,  using a small $t_0$, $t_0 = \Delta T = 6K$ updates (Section \ref{sssec:system}).  
Instead, we can increase $t_0$ and prune a better converged model, assuming that the weights will be better initialized for the pruning criterion (i.e., weight magnitude). However, we find simply increasing $t_0$ can only produce results similar to C2, since as discussed in Section \ref{ssec:block_wd}, weights learned during $(0, t_0)$ do not follow the $8\times1$ block structure, and the structured sparsity may prune out important weights in each block. 
 Therefore, next, we not only increase $t_0$ and also apply the additional group lasso weight decay during $(0, t_0 + n \Delta T)$. We find the resulting system C3.1  with $t_0 = 50K$ outperforms both baseline C1 and C2.

 Finally,  we increase $t_0 = 150K$ in system C3.2, and find (i) compared to the dense model B2 without any sparsity (Table \ref{tab:result3}),  C3.2 can match the topline non-streaming performance, and (ii) compared to baseline C1, C3.2 can effectively leverage self-supervised learning and provide a significantly improved sparse streaming model, by 11-16\% WER reductions.

%   encourages the small weights close to 0 in each block,   such two-stage pretraining is very effective that 
% and we expect the results in C3 to further improve for both modes. Also, currents results in C3 are based on pretraining with 15s segment
% and work in progress includes pretraining on 30s, as evidenced by comparing B2 and B3. Note that pretraining on 30s cannot converge without the auxiliary wav2vec criterion (as introduced in post), similar to the auxiliary RNN-T criterion \cite{liu2021improving}. Finally, we will perform IPL on top of C3, where it uses the performant non-streaming mode for pseudo-labeling. 
% ---------------------------------------------------------------------------------------------------------------------------------------------
% ---------------------------------------------------------------------------------------------------------------------------------------------
% \section{Related Work}
% \label{sec:related}

% ---------------------------------------------------------------------------------------------------------------------------------------------
\section{Conclusions}

Overall, we first present a dynamic Emformer segment sampling framework to enable a dual-mode encoder. We demonstrate that, jointly learning a featherweight sparse streaming ASR model and a large dense non-streaming model - in a single supernet - can provide competitive accuracies compared to learning each individually. 
Second, the proposed dual-mode encoder can dynamically use the non-streaming mode during the wav2vec 2.0 pre-training and perform dual-mode ASR thereafter, which allows for self-supervised learning equally helpful for the non-streaming mode and also to substantially improve the streaming ASR. 

Next, we show that the proposed group lasso weight decay can effectively address the block patterns as required in structured pruning, such that the self-supervised pre-training is able to identify a performant and robust sub-model for the downstream task. 
Finally, we conclude that for both self-supervised and supervised learning, the proposed supernet training of a sparse sub-model and a dense model jointly can provide an equally competitive non-streaming ASR model and also provide a noticeably improved sparse streaming model.

\bibliographystyle{IEEEbib}
\bibliography{thesis,refs}

\end{document}